\documentclass[11pt]{article}%
\usepackage{amsfonts}
\usepackage{amsmath}
\usepackage{amssymb}
\usepackage{graphicx}%
\newtheorem{theorem}{Theorem}

\newtheorem{corollary}[theorem]{Corollary}

\newtheorem{lemma}[theorem]{Lemma}

\newtheorem{proposition}[theorem]{Proposition}

\newenvironment{proof}[1][Proof]{\noindent\textbf{#1.} }{\ \rule{0.5em}{0.5em}}
\begin{document}

\title{The Multi-Dimensional Hardy Uncertainty Principle and its Interpretation in
Terms of the Wigner Distribution; Relation With the Notion of Symplectic Capacity}
\author{Maurice de Gosson\\\textit{Max-Planck-Institut f\"{u}r Mathematik }\\\textit{Pf. 7280, DE-53072 Bonn}
\and Franz Luef\thanks{This author has been supported by the \textit{European Union
EUCETIFA grant MEXT-CT-2004-517154.}}\\\textit{Universit\"{a}t Wien}\\\textit{Fakult\"{a}t f\"{u}r Mathematik, }\\\textit{Nordbergstrasse 15, AT-1090 Wien}}
\maketitle

\begin{abstract}
We extend Hardy's uncertainty principle \ for a square integrable function
$\psi$ and its Fourier transform to the $n$-dimensional case using a
symplectic diagonalization. We use this extension to show that Hardy's
uncertainty principle is equivalent to a statement on the Wigner distribution
$W\psi$ of $\psi$. We give a geometric interpretation of our results in terms
of the notion of symplectic capacity of an ellipsoid. Furthermore, we show
that Hardy's uncertainty principle is valid for a general Lagrangian frame of
the phase space. Finally, we discuss an extension of Hardy's theorem for the
Wigner distribution for exponentials with convex exponents.

\end{abstract}

\section{Introduction}

A folk metatheorem is that a function $\psi$ and its Fourier transform $F\psi$
cannot be simultaneously sharply localized. An obvious manifestation of this
\textquotedblleft principle\textquotedblright\ is when $\psi$ is of compact
support: in this case the Fourier transform $F\psi$ can be extended into an
entire function, and is hence never of compact support. A less trivial way to
express this kind of trade-off between $\psi$ and $F\psi$ was discovered in
1933 by G.H.\ Hardy \cite{ha33}. Hardy showed, using methods from complex
analysis (the Phragm\'{e}n--Lindel\"{o}f principle), that if $\psi\in
L^{2}(\mathbb{R})$ and its Fourier transform
\[
F\psi(p)=\frac{1}{\sqrt{2\pi\hbar}}\int_{-\infty}^{\infty}e^{-\tfrac{i}{\hbar
}px}\psi(x)dx
\]
satisfy, for $|x|+|p|\rightarrow\infty$, estimates
\begin{equation}
\psi(x)=\mathcal{O}(e^{-\tfrac{a}{2\hbar}x^{2}})\text{ \ , \ }F\psi
(p)=\mathcal{O}(e^{-\tfrac{b}{2\hbar}p^{2}}) \label{ha}%
\end{equation}
with $a,b>0$, then the following holds true:

\begin{itemize}
\item[(1)] \textit{If }$ab>1$\textit{ then }$\psi=0$\textit{;}

\item[(2)] \textit{If }$ab=1$\textit{ we have }$\psi(x)=Ce^{-\frac{a}{2\hbar
}x^{2}}$\textit{ for some complex constant }$C$\textit{;}

\item[(3)] \textit{If }$ab<1$\textit{ we have }$\psi(x)=Q(x)e^{-\frac
{a}{2\hbar}x^{2}}$\textit{ where }$Q$\textit{ is a polynomial function.}
\end{itemize}

Recently, researchers in harmonic analysis and time-frequency analysis have
formulated variants of Hardy's theorem for phase space representations
(time-frequency representations) such as the Wigner distribution, see
\cite{bodeja03,grzi01,gr03-2}. The results in \cite{grzi01,gr03-2} are deduced
from Hardy's theorem for a carefully chosen square-integrable function and its
Fourier transform. In \cite{bodeja03} a multidimensional extension of Hardy's
theorem is presented, which are based on an extension of the
Phragm\'{e}n--Lindel\"{o}f principle to several complex variables. The results
of \cite{bodeja03} have in a sense the same flavor as our statements, but they
are of a completely different nature. Actually, we only invoke real variable
methods in our proof of the $n$-dimensional Hardy theorem.

The principal aim of this paper is to reformulate Hardy's theorem in terms of
phase-space objects. We will actually give a non-trivial restatement of
Hardy's theorem for functions $\psi\in L^{2}(\mathbb{R}^{n})$ satisfying
estimates%
\begin{equation}
\psi(x)=\mathcal{O}(e^{-\tfrac{1}{2\hbar}x^{T}Ax})\text{ \ , \ }%
F\psi(p)=\mathcal{O}(e^{-\tfrac{1}{2\hbar}x^{T}Bx}) \label{hb}%
\end{equation}
where $A,B$ are positive-definite symmetric matrices, and show that the
estimates
\begin{equation}
\psi(x)=\mathcal{O(}e^{-\tfrac{1}{2\hbar}Ax^{2}})\text{ \ , \ }F\psi
(p)=\mathcal{O}(e^{-\tfrac{1}{2\hbar}Bx^{2}}) \label{hc}%
\end{equation}
are equivalent to a single estimate
\[
W\psi(x,p)=\mathcal{O(}e^{-\tfrac{1}{\hbar}(x^{T}Ax+p^{T}Bp)})\text{ \ for
\ }|x|+|p|\rightarrow\infty\text{ }%
\]
for the Wigner transform of $\psi$. This theorem provides a positive answer to
a question raised by Gr\"ochenig in \cite{gr03-2} on the equivalence of
uncertainty principles for a function and its Fourier transform and
uncertainty principles for the Wigner distribution (or more generally, for any
phase space representation).

We will see that the geometric interpretation of the conditions on the
matrices $A,B$ is that the symplectic capacity of the \textquotedblleft Wigner
ellipsoid\textquotedblright\
\[
\mathcal{W}:x^{T}Ax+p^{T}Bp\leq\hbar
\]
is at least $\frac{1}{2}h$, the half of the quantum of action. This property
is related to the fact that the notion of symplectic capacity is a natural
tool for expressing the uncertainty principle of quantum mechanics in a
symplectically covariant and intrinsic form as discussed in de Gosson
\cite{go05,Birk,go06}; also see de Gosson and Luef \cite{GosLue,Goluahp}); it
turns out that, more generally, if a function $\psi\in L^{2}(\mathbb{R}^{n})$
satisfies an estimate%
\[
W\psi(z)=\mathcal{O(}e^{-\tfrac{1}{\hbar}z^{T}Mz})\text{ \ for \ }%
|z|\rightarrow\infty\text{ }%
\]
where $z=(x,p)$ then the symplectic capacity of $\mathcal{W}:z^{T}Mz\leq\hbar$
must be $\geq\frac{1}{2}h.$

Actually, we also state a version of Hardy's theorem, which is valid for an
arbitrary pair of Lagrangian frames, i.e a transversal pair of Lagrangian
planes. Therefore, the main results of our investigation provides a rigorous
justification of a reformulation of the uncertainty principle in \cite{gust82}
due to Guillemin and Sternberg: \textquotedblleft The smallest subsets of
classical phase space in which the presence of quantum particle can be
detected are its Lagrangian submanifolds.\textquotedblright\ Consequently, one
could say that one of the main aims of the present article is to exploit the
symplectic nature of Hardy's uncertainty principle in the sense of Guillemin
and Sternberg.

Our work is structured as follows:

\begin{itemize}
\item In Section \ref{sec2} we prove a multi-dimensional variant of Hardy's
theorem, as a property of the symplectic spectrum of the matrix $%
\begin{pmatrix}
A & 0\\
0 & B
\end{pmatrix}
$ extracted from the conditions (\ref{hc}) where $A$ and $B$ are
positive-definite symmetric matrices; here $F$ denotes the $\hbar$-dependent
$n$-dimensional Fourier transform. In Lemma \ref{lem1} (Subsection
\ref{subs1}) we show that it is possible to perform a \emph{symplectic}
diagonalization of a positive-definite block-diagonal matrix using symplectic
block-diagonal matrices.

\item In Section \ref{sec2bis} we give a purely geometric interpretation of
Hardy's uncertainty principle in terms of the notion of the symplectic
capacity, which is closed related to Gromov's non-squeezing theorem; we take
the opportunity to quickly review the main definitions and properties
concerning these objects. In particular, we point out that all symplectic
capacities agree on phase-space ellipsoids;

\item In Section \ref{sec3} we restate the results above in terms of the
Wigner transform, by showing that the conditions (\ref{hc}) are equivalent to
\begin{equation}
W\psi(x,p)\leq Ce^{-\frac{1}{\hslash}(x^{T}Ax+p^{T}Bp)} \label{hd}%
\end{equation}
for some constant $C\geq0$. In Subsection \ref{sec4} we give an equivalent
geometric statement of the results above in terms of the topological notion of
symplectic capacity. We will in fact prove that if (\ref{ha}) holds for
$\psi\neq0$ then the symplectic capacity of the phase space
ellipsoid\ $\mathcal{W}:\frac{1}{2}z^{T}\Sigma^{-1}z\leq1$ is at least
$\frac{1}{2}h$; here $\Sigma$ is the covariance matrix defined by%
\[
\Sigma=%
\begin{pmatrix}
(\Delta x)^{2} & 0\\
0 & (\Delta p)^{2}%
\end{pmatrix}
\]
where $\Delta x=\sqrt{\hbar/2a}$ and $\Delta p=\sqrt{\hbar/2b}$. The condition
$ab\leq1$ is thus equivalent to the Heisenberg inequality $\Delta x\Delta
p\geq\frac{1}{2}\hbar$.

\item In Section \ref{seclast} we give two non-obvious extensions of the
results obtained in the previous sections. The first extension (Subsection
\ref{seclast1}) consists in replacing the $x,p$ coordinate system by an
arbitrary \textquotedblleft Lagrangian frame\textquotedblright\ $(\ell
,\ell^{\prime})$ and to use the transitivity of the action of the symplectic
group on the set of all such frames. In the second extension (Subsection
\ref{seclast2}) we consider estimates of the type $W\psi(z)\leq Ce^{-\frac
{1}{\hbar}Q(z)}$ where $Q$ is a twice continuously differentiable function
which is uniformly convex. We express the necessary condition on that function
in terms of the symplectic capacity of the convex set $Q(z)\leq\hbar.$
\bigskip
\end{itemize}

\noindent\textbf{Notation}. We will use the shorthand notation $z=(x,p)$ for
points of the phase space $\mathbb{R}^{2n}\equiv\mathbb{R}^{n}\times
\mathbb{R}^{n}$. The symplectic product of two vectors $z=(x,p)$, $z^{\prime
}=(x^{\prime},p^{\prime})$ in $\mathbb{R}^{2n}$ is
\[
\sigma(z,z^{\prime})=p\cdot x-p^{\prime}\cdot x
\]
where the dot $\cdot$ stands for the usual (Euclidean) scalar product;
alternatively $\sigma(z,z^{\prime})=Jz\cdot z^{\prime}$ where $J=%
\begin{pmatrix}
0 & I\\
-I & 0
\end{pmatrix}
$ ($0$ (resp. $I$) is the zero (resp. identity) matrix of order $n$). The
symplectic group is denoted by $\operatorname{Sp}(n)$: we have $S\in
\operatorname{Sp}(n)$ if and only if $S$ is a real matrix of order $2n$ such
that $\sigma(Sz,Sz^{\prime})=\sigma(z,z^{\prime})$; equivalently
$S^{T}JS=SJS^{T}=J$.

When $M$ is a symmetric matrix we will often write $Mx^{2}$, $Mp^{2},$
$Mz^{2}$ instead of $Mx\cdot x$ (or $x^{T}Mx$), $Mp\cdot p$, $Mz\cdot z.$ To
express that $M$ is symmetric and positive-definite we will use the notation
$M>0$.

$F$ denotes the $n$-dimensional $\hbar$-dependent Fourier transform. It is the
unitary operator $L^{2}(\mathbb{R}^{n})\longrightarrow L^{2}(\mathbb{R}^{n})$
defined for $\psi\in\mathcal{S}(\mathbb{R}^{n})$ by%
\begin{equation}
F\psi(p)=\left(  \tfrac{1}{2\pi\hbar}\right)  ^{n/2}\int e^{-\tfrac{i}{\hbar
}p\cdot x}\psi(x)d^{n}x\text{.} \label{FT}%
\end{equation}

\section{Hardy's Theorem in Dimension $n$\label{sec2}}

Using classical results on the simultaneous diagonalization of a pair of
symmetric matrices it is possible to extend Hardy's theorem to the case of
$\mathbb{R}^{n}$ (see for instance Sitaram \textit{et al}. \cite{sisu}). We
are going to prove a variant of this result using a \emph{symplectic}
diagonalization; this will allow us to relate our statements to the notion of
symplectic capacity later on in this work.

\subsection{A symplectic diagonalization result\label{subs1}}

The following result, although being of an elementary nature is very useful.
We will see that it is a refined version of Williamson's diagonalization
theorem \cite{wi36} in the block-diagonal case.

We make the preliminary observation that if $A$ and $B$ are positive definite
matrices then the eigenvalues of $AB$ are real because $AB$ has the same
eigenvalues as the symmetric matrix $A^{1/2}BA^{1/2}$.

\begin{lemma}
\label{lem1}Let $A$, $B>0.$ There exists $L\in GL(n,\mathbb{R})$ such that%
\begin{equation}
L^{T}AL=L^{-1}B(L^{T})^{-1}=\Lambda\label{lalb}%
\end{equation}
\ where $\Lambda=\operatorname*{diag}(\sqrt{\lambda_{1}},...,\sqrt{\lambda
_{n}})$ is the diagonal matrix whose eigenvalues are the square roots of the
eigenvalues $\lambda_{1},...,\lambda_{n}$ of $AB$.
\end{lemma}

\begin{proof}
We claim that there exists $R\in GL(n,\mathbb{R})$ such that%
\begin{equation}
R^{T}AR=I\text{ and }R^{-1}B(R^{T})^{-1}=D \label{diag1}%
\end{equation}
where $D=\operatorname*{diag}(\lambda_{1},...,\lambda_{n})$. In fact, first
choose $P\in GL(n,\mathbb{R})$ such that $P^{T}AP=I$ and set $B_{1}^{-1}%
=P^{T}B^{-1}P$. Since $B_{1}^{-1}$ is symmetric, there exists $H\in
O(n,\mathbb{R})$ such that $B_{1}^{-1}=H^{T}D^{-1}H$ where $D^{-1}$ is
diagonal. Set now $R=PH^{T}$; we have $R^{T}AR=I$ and also
\[
R^{-1}B(R^{T})^{-1}=HP^{-1}B(P^{T})^{-1}H^{T}=HB_{1}H^{T}=D
\]
hence the equalities (\ref{diag1}). Let $\Lambda=\operatorname*{diag}%
(\sqrt{\lambda_{1}},...,\sqrt{\lambda_{n}})$. Since
\[
R^{T}AB(R^{T})^{-1}=R^{T}AR(R^{-1}B(R^{T})^{-1})=D
\]
the diagonal elements of $D$ are indeed the eigenvalues of $AB$ hence
$D=\Lambda^{2}$. Setting $L=R\Lambda^{1/2}$ we have
\begin{gather*}
L^{T}AL=\Lambda^{1/2}R^{T}AR\Lambda^{1/2}=\Lambda\\
L^{-1}B(L^{-1})^{T}=\Lambda^{-1/2}R^{-1}B(R^{T})^{-1}\Lambda^{-1/2}=\Lambda
\end{gather*}
hence our claim.
\end{proof}

The result above is a precise statement of a classical theorem of Williamson
\cite{wi36} in the block-diagonal case. That theorem says that every
positive-definite symmetric matrix can be diagonalized using symplectic
matrices. More precisely: let $M$ be a positive definite real $2n\times2n$
matrix; the eigenvalues of $JM$ are those of the antisymmetric matrix
$M^{1/2}JM^{1/2}$ and are thus of the type $\pm i\lambda_{j}^{\sigma}$ with
$\lambda_{j}^{\sigma}>0$. We have:

\begin{theorem}
[Williamson](i) There exists $S\in\operatorname*{Sp}(n)$ such that $S^{T}MS=%
\begin{pmatrix}
\Lambda & 0\\
0 & \Lambda
\end{pmatrix}
$ where $\Lambda=\operatorname*{diag}(\lambda_{1}^{\sigma},...,\lambda
_{n}^{\sigma})$.

(ii) The symplectic matrix $S$ is unique up to a unitary factor: if
$S^{\prime}$ is another Williamson diagonalizing symplectic matrix then
$S(S^{\prime})^{-1}\in U(n)$.
\end{theorem}

\begin{proof}
(i) See for instance \cite{Birk,hoze94} for \textquotedblleft
modern\textquotedblright\ proofs. (ii) See de Gosson \cite{Birk}.
\end{proof}

We will always arrange the $\lambda_{j}^{\sigma}$ in decreasing order:
$\lambda_{1}^{\sigma}\geq\lambda_{2}^{\sigma}\geq\cdot\cdot\cdot\geq
\lambda_{n}^{\sigma}$ and call $(\lambda_{1}^{\sigma},...,\lambda_{n}^{\sigma
})$ the \textit{symplectic spectrum} of the positive definite matrix $M$. The
positive numbers $\lambda_{j}^{\sigma}$ (which only depend on $M$, and not on
$S$) are the \textit{Williamson invariants} of $M$. Writing the diagonalizing
symplectic matrix as $S=(X_{1},...,X_{n};Y_{1},...,Y_{n})$ where the $X_{j}$
and $Y_{k}$ are column vectors, the set $\mathcal{B}=\{X_{1},...,X_{n}%
;Y_{1},...,Y_{n}\}$ is called a \textit{Williamson basis} for $M$ (it is of
course not uniquely defined in general). A Williamson basis is a symplectic
basis of $(\mathbb{R}^{2n},\sigma)$, that is $\sigma(X_{j},X_{k})=\sigma
(X_{j},X_{k})$ and $\sigma(Y_{j},X_{k})=\delta_{jk}$ for $1\leq j,k\leq n$.

The following result relates Lemma \ref{lem1} to Williamson's theorem:

\begin{lemma}
\label{lem2} Let $A,B>0$. The symplectic spectrum $(\lambda_{1}^{\sigma
},...,\lambda_{n}^{\sigma})$ of $M=%
\begin{pmatrix}
A & 0\\
0 & B
\end{pmatrix}
$ consists of the decreasing sequence $\sqrt{\lambda_{1}}\geq\cdot\cdot
\cdot\geq\sqrt{\lambda_{n}}$ of square roots of the eigenvalues $\lambda_{j}$
of $AB.$
\end{lemma}

\begin{proof}
Let $(\lambda_{1}^{\sigma},...,\lambda_{n}^{\sigma})$ be the symplectic
spectrum of $M$. The $\lambda_{j}^{\sigma}$ are the eigenvalues of
\[
JM=%
\begin{pmatrix}
0 & B\\
-A & 0
\end{pmatrix}
;
\]
they are thus the moduli of the zeroes of the polynomial%
\[
P(t)=\det(t^{2}I+AB)=\det(t^{2}I+D)
\]
where $D=\operatorname*{diag}(\lambda_{1},...,\lambda_{n})$; these zeroes are
the numbers $\pm i\sqrt{\lambda_{j}},$ $j=1,...,n$; the result follows.
\end{proof}

For $L$ invertible set
\begin{equation}
M_{L}=%
\begin{pmatrix}
L^{-1} & 0\\
0 & L^{T}%
\end{pmatrix}
. \label{ml}%
\end{equation}
Obviously $M_{L}\in\operatorname*{Sp}(n)$; Lemma \ref{lem1} can be restated by
saying that if $(A,B)$ is a pair of symmetric positive definite then there
exists $L$ such that
\begin{equation}%
\begin{pmatrix}
A & 0\\
0 & B
\end{pmatrix}
=M_{L^{T}}%
\begin{pmatrix}
\Lambda & 0\\
0 & \Lambda
\end{pmatrix}
M_{L}. \label{diag}%
\end{equation}
Lemma \ref{lem1} is thus a precise version of Williamson's theorem for
block-diagonal positive matrices ---it is not at all obvious from the
statement of this theorem that such a matrix can be diagonalized using only a
block-diagonal symplectic matrix!

\subsection{Application to Hardy's theorem}

Lemma \ref{lem1} allows us to give a simple proof of a multi-dimensional
version of this theorem. The following elementary remark will be useful:

\begin{lemma}
\label{lemtens}Let $n>1$. For $1\leq j\leq n$ let $f_{j}$ be a function of
$(x_{1},..,\widetilde{x_{j}},...,x_{n})\in\mathbb{R}^{n-1}$ (the tilde
$\widetilde{}$ suppressing the term it covers), and $g_{j}$ a function of
$x_{j}\in\mathbb{R}$. If
\[
h=f_{1}\otimes g_{1}=\cdot\cdot\cdot=f_{n}\otimes g_{n}%
\]
then there exists a constant $C$ such that $h=C(g_{1}\otimes\cdot\cdot
\cdot\otimes g_{n}).$
\end{lemma}

\begin{proof}
Assume that $n=2$; then
\[
h(x_{1},x_{2})=f_{1}(x_{2})g_{1}(x_{1})=f_{2}(x_{1})g_{2}(x_{2}).
\]
If $g_{1}(x_{1})g_{2}(x_{2})\neq0$ then
\[
f_{1}(x_{2})/g_{2}(x_{2})=f_{2}(x_{1})/g_{1}(x_{1})=C
\]
hence $f_{1}(x_{2})=Cg_{2}(x_{2})$ and $h(x_{1},x_{2})=Cg_{1}(x_{1}%
)g_{2}(x_{2})$. If $g_{1}(x_{1})g_{2}(x_{2})=0$ then $h(x_{1},x_{2})=0$ hence
$h(x_{1},x_{2})=Cg_{1}(x_{1})g_{2}(x_{2})$ in all cases. The general case
follows by induction on the dimension $n$: suppose that
\[
h=f_{1}\otimes g_{1}=\cdot\cdot\cdot=f_{n}\otimes g_{n}=f_{n+1}\otimes
g_{n+1};
\]
for fixed $x_{n+1}$ the function $k=f_{1}\otimes g_{1}=\cdot\cdot\cdot
=f_{n}\otimes g_{n}$ is given by
\[
k(x,x_{n+1})=C(x_{n+1})g_{1}(x_{1})\cdot\cdot\cdot g_{n}(x_{n}).
\]
Since we also have
\[
k(x,x_{n+1})=f_{n+1}(x_{1},...,x_{n})g_{n+1}(x_{n+1})
\]
it follows that $C(x_{n+1})=C$.
\end{proof}

\begin{theorem}
\label{prop1}Let $A$ and $B$ be two real positive definite matrices and
$\psi\in L^{2}(\mathbb{R}^{n})$, $\psi\neq0$. Assume that
\begin{equation}
|\psi(x)|\leq C_{A}e^{-\tfrac{1}{2\hbar}Ax^{2}}\text{ \ and \ }|F\psi(p)|\leq
C_{B}e^{-\tfrac{1}{2\hbar}Bp^{2}} \label{AB}%
\end{equation}
for some constants $C_{A},C_{B}>0$. Then:

(i) The eigenvalues $\lambda_{j}$, $j=1,...,n$, of $AB$ are $\leq1$;

(ii) If $\lambda_{j}=1$ for all $j$, then $\psi(x)=Ce^{-\frac{1}{2\hbar}%
Ax^{2}}$ for some some complex constant $C$.

(iii) If $\lambda_{j}<1$ for some $j$ then $\psi(x)=Q(x)e^{-\tfrac{1}{2\hbar
}Ax^{2}}$ for some polynomial function $Q:\mathbb{R}^{n}\longrightarrow
\mathbb{C}$.
\end{theorem}

\begin{proof}
\textit{Proof of (i)}. It is of course no restriction to assume that
$C_{A}=C_{B}=C$. Let $L$ be as in Lemma \ref{lem1} and order the eigenvalues
of $AB$ decreasingly: $\lambda_{1}\geq\lambda_{2}\geq\cdot\cdot\cdot
\geq\lambda_{n}$. It suffices to show that $\lambda_{1}\leq1$. Setting
$\psi_{L}(x)=\psi(Lx)$ we have
\[
F\psi_{L}(p)=F\psi((L^{T})^{-1}p);
\]
in view of (\ref{lalb}) in Lemma \ref{lem1} condition (\ref{AB}) is equivalent
to%
\begin{equation}
|\psi_{L}(x)|\leq Ce^{-\tfrac{1}{2\hbar}\Lambda x^{2}}\text{ \ \textit{and}
\ }|F\psi_{L}(p)|\leq Ce^{-\tfrac{1}{2\hbar}\Lambda p^{2}} \label{psil}%
\end{equation}
where $\Lambda=\operatorname*{diag}(\lambda_{1},\lambda_{2},...,\lambda_{n})$.
Setting $\psi_{L,1}(x_{1})=\psi_{L}(x_{1},0,...,0)$ we have%
\begin{equation}
|\psi_{L,1}(x_{1})|\leq Ce^{-\tfrac{1}{2\hbar}\lambda_{1}x_{1}^{2}}.
\label{psil1}%
\end{equation}
On the other hand, by the Fourier inversion formula,
\begin{align*}
\int F\psi_{L}(p)dp_{2}\cdot\cdot\cdot dp_{n}  &  =(2\pi\hbar)^{n/2}\iint
e^{-\frac{i}{\hbar}p\cdot x}\psi_{L}(x)dxdp_{2}\cdot\cdot\cdot dp_{n}\\
&  =(2\pi\hbar)^{(n-1)/2}F\psi_{L,1}(p_{1})
\end{align*}
and hence%
\begin{equation}
|F\psi_{L,1}(p_{1})|\leq C_{L,1}e^{-\tfrac{1}{2\hbar}\lambda_{1}p_{1}^{2}}
\label{psil2}%
\end{equation}
for some constant $C_{L,1}>0$. Applying Hardy's theorem to the inequalities
(\ref{psil1}) and (\ref{psil2}) we must have $\lambda_{1}^{2}\leq1$ hence the
assertion (i). \textit{Proof of (ii).} The condition $\lambda_{j}=1$ for all
$j$ means that%
\begin{equation}
|\psi_{L}(x)|\leq Ce^{-\tfrac{1}{2\hbar}x^{2}}\text{ \ \textit{and} \ }%
|F\psi_{L}(p)|\leq Ce^{-\tfrac{1}{2\hbar}p^{2}} \label{psi3}%
\end{equation}
for some $C>0$. Let us keep $x^{\prime}=(x_{2},...,x_{n})$ constant; the
partial Fourier transform of $\psi_{L}$ in the $x_{1}$ variable is $F_{1}%
\psi_{L}=(F^{\prime})^{-1}F\psi_{L}$ where $(F^{\prime})^{-1}$ is the inverse
Fourier transform in the $x^{\prime}$ variables, hence there exists
$C^{\prime}>0$ such that%
\[
|F_{1}\psi_{L}(x_{1},x^{\prime})|\leq\left(  \tfrac{1}{2\pi\hbar}\right)
^{\frac{n-1}{2}}\int|F\psi_{L}(p)|dp_{2}\cdot\cdot\cdot dp_{n}\leq C^{\prime
}e^{-\tfrac{1}{2\hbar}p_{1}^{2}}.
\]
Since $|\psi_{L}(x)|\leq C(x^{\prime})e^{-\frac{1}{2\hbar}x_{1}^{2}}$ with
$C(x^{\prime})\leq e^{-\frac{1}{2\hbar}x^{\prime2}}$ it follows from Hardy's
theorem that we can write%
\[
\psi_{L}(x)=f_{1}(x^{\prime})e^{-\tfrac{1}{2\hbar}x_{1}^{2}}%
\]
for some real $C^{\infty}$ function $f_{1}$ on $\mathbb{R}^{n-1}$. Applying
the same argument to the remaining variables $x_{2},...,x_{n}$ we conclude
that there exist $C^{\infty}$ functions $f_{j}$ for $j=2,...,n$, such that
\begin{equation}
\psi_{L}(x)=f_{j}(x_{1},..,\widetilde{x_{j}},...,x_{n})e^{-\tfrac{1}{2\hbar
}x_{1}^{2}}. \label{fj}%
\end{equation}
In view of Lemma \ref{lemtens} above we have $\psi_{L}(x)=C_{L}e^{-\frac
{1}{2\hbar}x^{2}}$ for some constant $C_{L}$; since $\Lambda=I=L^{T}AL$ we
thus have $\psi(x)=C_{L}e^{-Ax^{2}/2\hbar}$ as claimed. \textit{Proof of
(iii). Assume that }$\lambda_{1}<1$ for $j\in\mathcal{J}$, $\mathcal{J}$ a
subset of $\{1,...,n\}$. By the same argument as in the proof of part (ii)
establishing formula (\ref{fj}), we infer, using Hardy's theorem in the case
$ab<1$, that%
\[
\psi_{L}(x)=f_{j}(x_{1},..,\widetilde{x_{j}},...,x_{n})Q_{j}(x_{j}%
)e^{-\tfrac{1}{2\hbar}x_{j}^{2}}%
\]
where $Q_{j}$ is a polynomial with degree $0$ if $j\notin\mathcal{J}$. One
concludes the proof using one again Lemma \ref{lemtens}.
\end{proof}

\section{Geometric interpretation\label{sec2bis}}

Let us give a geometric interpretation of Theorem \ref{prop1}. We begin by
making an obvious observation: Hardy's uncertainty principle can be restated
by saying that if $\psi\neq0$ then the conditions $\psi(x)=\mathcal{O}%
(e^{-\frac{1}{2\hbar}ax^{2}})$\ \textit{and} \ $F\psi(p)=\mathcal{O}%
(e^{-\frac{1}{2\hbar}bp^{2}})$ imply that the ellipse\ $\mathcal{W}%
:ax^{2}+bp^{2}\leq\hbar$ has area $\mathcal{\pi\hbar}/\sqrt{ab}\geq
\mathcal{\pi\hbar=}\frac{1}{2}h$:
\[
\operatorname*{Area}(\mathcal{W)\geq}\frac{1}{2}h\text{.}%
\]
More precisely:

\emph{If the area of the ellipse }$\mathcal{W}$\emph{ is smaller than }%
$\frac{1}{2}h$\emph{ then }$\psi=0$\emph{; if this area equals }$\frac{1}{2}%
h$\emph{ then }$\psi(x)=Ce^{-\frac{1}{2\hbar}ax^{2}}$\emph{ and if it is
larger than }$\frac{1}{2}h$\emph{ then }$\psi(x)=Q(x)e^{-\frac{1}{2\hbar
}ax^{2}}$\emph{ where }$\emph{Q}$\emph{ is a polynomial function.}

When trying to generalize this observation to higher dimensions, one should
resist the pitfall of copying the statement above \textit{mutatis mutandis}
and replacing everywhere the word \textquotedblleft area\textquotedblright\ by
\textquotedblleft volume\textquotedblright. As we will see, volume is not the
right answer; one has instead to use the more subtle notion of
\textit{symplectic capacity}, introduced by Ekeland and Hofer \cite{EH}
following Gromov's work \cite{gr85} on pseudoholomorphic curves.

\subsection{Symplectic capacities and symplectic spectrum}

A symplectic capacity on the symplectic space $(\mathbb{R}^{2n},\sigma)$
assigns to every subset $\Omega$ of $\mathbb{R}^{2n}$ a number $c(\Omega
)\geq0$ or $+\infty$; this assignment has the four properties listed below.
(We denote by\textit{ }$B(R)$ the ball $|z|\leq R$\ and by $Z_{j}(R)$ the
cylinder $x_{j}^{2}+p_{j}^{2}\leq R^{2}$.)

\begin{description}
\item[SC1] \textbf{Monotonicity}:\textit{\ }$c(\Omega)\leq c(\Omega^{\prime})$
if $\Omega\subset\Omega^{\prime}$;

\item[SC2] \textbf{Symplectic invariance}:\textit{\ }$c(f(\Omega))=c(\Omega)$
\textit{for every symplectomorphism }$f$\textit{\ defined near }$\Omega$;

\item[SC3] \textbf{Conformality}:\textit{\ }$c(\lambda\Omega)=\lambda
^{2}c(\Omega)$ \textit{if} $\lambda\in\mathbb{R}$\textit{;}

\item[SC4] \textbf{Nontriviality}:\textit{\ We have} $c(B(R))=c(Z_{j}(R))=\pi
R^{2}$.
\end{description}

A fundamental example of symplectic capacity is provided by the
\textquotedblleft Gromov width\textquotedblright, defined by
\begin{equation}
c_{\text{Gr}}(\Omega)=\sup_{f\in\operatorname*{Symp}(n)}\{\pi r^{2}%
:f(B(R))\subset\Omega\} \label{cgr}%
\end{equation}
where $\operatorname*{Symp}(n)$ is the group of all symplectomorphisms of
$(\mathbb{R}^{2n},\sigma)$. Properties \textit{(i)}--\textit{(iii)} and
$c_{\text{Gr}}(B(R))=\pi R^{2}$ are trivially verified; that we also have
$c_{\text{Gr}}(Z_{j}(R))=\pi R^{2}$ is just Gromov's non-squeezing theorem
\cite{gr85} which asserts that a phase-space ball cannot be squeezed inside a
symplectic cylinder with smaller radius using symplectomorphisms (but such a
squeezing can, of course, be performed using general volume-preserving diffeomorphisms).

We will also use the \textquotedblleft linear symplectic
capacity\textquotedblright\ $c_{\text{lin}}$ defined by
\begin{equation}
c_{\text{lin}}(\Omega)=\sup_{f\in\operatorname*{ISp}(n)}\{\pi r^{2}%
:f(B(R))\subset\Omega\} \label{lincap}%
\end{equation}
where $f$ this time ranges over the group $\operatorname*{ISp}(n)$ of all
affine symplectic automorphisms of $(\mathbb{R}^{2n},\sigma)$ (the
\textquotedblleft inhomogeneous symplectic group\textquotedblright). The
capacity $c_{\text{lin}}$ has the same properties as general symplectic
capacities, except that it is only invariant under linear or affine symplectomorphisms.

We have the following result, which allows us to talk about \emph{the}
symplectic capacity of a phase-space ellipsoid:

\begin{lemma}
\label{lem3}For $M>0$ let $\Omega_{M}=\{z\in\mathbb{R}^{2n}:Mz^{2}\leq1\}$.
For any symplectic capacity $c$ on $(\mathbb{R}^{2n},\sigma)$ we have%
\begin{equation}
c(\Omega M)=c_{\text{lin}}(\Omega_{M})=\frac{\pi}{\lambda_{1}^{\sigma}}
\label{capell}%
\end{equation}
where $\lambda_{1}^{\sigma}\geq\cdot\cdot\cdot\geq\lambda_{n}^{\sigma}$ is the
symplectic spectrum of $M$.
\end{lemma}

\begin{proof}
See for instance Hofer and Zehnder \cite{hoze94}, Proposition 2, \S 2.1, p. 54
or de Gosson \cite{Birk}, Proposition 8.25, p. 251 (where $\lambda_{n}%
^{\sigma}$ should be replaced by $\lambda_{1}^{\sigma}$).
\end{proof}

\subsection{Application to the Wigner ellipsoid}

We can restate Hardy's theorem in a very simple geometric way in terms of the
symplectic capacity of the \textquotedblleft Wigner
ellipsoid\textquotedblright\ (the terminology seems to be due to Littlejohn
\cite{Littlejohn}):\ 

\begin{proposition}
\label{prop2}Let $\psi\in L^{2}(\mathbb{R}^{n})$, $\psi\neq0$. Assume that
\begin{equation}
|\psi(x)|\leq C_{A}e^{-\tfrac{1}{2\hbar}Ax^{2}}\text{ \ and \ }|F\psi(p)|\leq
C_{B}e^{-\tfrac{1}{2\hbar}Bp^{2}}. \label{psiab}%
\end{equation}
Then the symplectic capacity of the Wigner ellipsoid
\[
\mathcal{W}:Ax^{2}+Bp^{2}\leq\hbar
\]
satisfies $c(\mathcal{W})\geq\frac{1}{2}h$.
\end{proposition}

\begin{proof}
Setting $M=%
\begin{pmatrix}
A & 0\\
0 & B
\end{pmatrix}
$ the equation of $\mathcal{W}$ is $Mz^{2}\leq\hbar$. Consider the ellipsoid
$\Omega_{M}:Mz^{2}\leq1$. Let $(\lambda_{1}^{\sigma},\lambda_{2}^{\sigma
},...,\lambda_{n}^{\sigma})$ be the symplectic spectrum of $M$; by formula
(\ref{capell}) in Lemma \ref{lem3} we have $c(\Omega_{M})=\pi/\lambda
_{1}^{\sigma}$. In view of Lemma \ref{lem2} $\lambda_{j}^{\sigma}%
=\sqrt{\lambda_{j}}$ where the $\lambda_{j}$ are the eigenvalues of $AB$, and
by Theorem \ref{prop1} we must have $\lambda_{j}\leq1$, hence $c(\Omega
_{M})\geq\pi$. Since $\mathcal{W}=\sqrt{\hbar}\Omega_{M}$ we have
$c(\mathcal{W})=\hbar c(\Omega_{M})$ in view of the conformality property
(CZ3); the result follows.
\end{proof}

The result above will be extended to the Wigner distribution in next section.

\section{Hardy's Theorem and Wigner's distribution\label{sec3}}

It turns out that Hardy's theorem -- which involves two conditions, one about
a function and the other about the Fourier transform of that function -- is
equivalent to a \textit{single} \textit{condition} on the \textit{Wigner
transform} of $\psi$. This condition will be made explicit in Theorem
\ref{propmain} below; let us first prove some preliminary results about Wigner transforms.

\subsection{Wigner Distributions}

The Wigner transform of a function was introduced by Wigner in \cite{Wigner},
following joint work with Szilard. It is defined, for $\psi\in L^{2}%
(\mathbb{R}^{n})$, by the formula%
\begin{equation}
W\psi(z)=\left(  \tfrac{1}{2\pi\hbar}\right)  ^{n}\int e^{-\tfrac{i}{\hbar
}p\cdot y}\psi(x+\tfrac{1}{2}y)\overline{\psi(x-\tfrac{1}{2}y)}d^{n}y.
\label{wpsi1}%
\end{equation}
The function $W\psi$ has a simple interpretation in terms on the theory of
Weyl pseudodifferential operators. For $\psi\in L^{2}(\mathbb{R}^{n})$ with
$||\psi||=1$ consider the orthogonal projection $P_{\psi}$ on the ray
$\{\lambda\psi:\lambda\in\mathbb{C}\}$; we have $P_{\psi}\phi(x)=(\phi
|\psi)\psi$ for $\phi\in L^{2}(\mathbb{R}^{n})$ hence the operator kernel of
$P_{\psi}$ is $K_{\psi}=\psi\otimes\overline{\psi}$. Writing $P_{\psi}$ in
Weyl operator form we have%
\[
P_{\psi}\phi(x)=\iint e^{\tfrac{i}{\hbar}p\cdot(x-y)}\rho_{\psi}(\tfrac{1}%
{2}(x+y),p)\phi(y)d^{n}pd^{n}y
\]
where the symbol $\rho_{\psi}$ is given by%
\[
\rho_{\psi}(x,p)=\left(  \tfrac{1}{2\pi\hbar}\right)  ^{n}K_{\psi}(x+\tfrac
{1}{2}y,x-\tfrac{1}{2}y)=W\psi(z).
\]

More generally, one might want to consider the cross-Wigner transform (also
called Wigner--Moyal transform) which associates to a pair $(\psi,\phi)\in
L^{2}(\mathbb{R}^{n})\times L^{2}(\mathbb{R}^{n})$ the function
\[
W(\psi,\phi)(z)=\left(  \tfrac{1}{2\pi\hbar}\right)  ^{n}\int e^{-\tfrac
{i}{\hbar}p\cdot y}\psi(x+\tfrac{1}{2}y)\overline{\phi(x-\tfrac{1}{2}y)}%
d^{n}y;
\]
(it is the Weyl symbol of the operator defined by the kernel $\psi
\otimes\overline{\phi)\text{; }}$of course $W(\psi,\phi)=W\psi$. The following
properties of the (cross-) Wigner transform are well-known

\begin{description}
\item[W1] $W(\psi,\phi)=\overline{W(\phi,\psi)}$ (hence $W\psi$ is real);

\item[W2] If $\psi,F\psi\in L^{1}(\mathbb{R}^{n})\cap L^{2}(\mathbb{R}^{n})$
then
\begin{equation}
\int W\psi(z)d^{n}p=|\psi(x)|^{2}\text{ , }\int W\psi(z)d^{n}x=|F\psi(p)|^{2}.
\label{wpsi2}%
\end{equation}

\end{description}

Recall that the metaplectic group $\operatorname*{Mp}(n)$ is generated by the
following unitary operators on $L^{2}(\mathbb{R}^{2n})$ (\cite{Birk,fo89}):
the scaling operators $\widehat{M}_{L,m}$ ($L\in GL(n,\mathbb{R})$), the
\textquotedblleft chirps\textquotedblright\ $\widehat{V}_{P}$ ($P=P^{T}$), and
the modified Fourier transform $\widehat{J}=i^{-n/2}F$; by definition
\begin{equation}
\widehat{M}_{L,m}\psi(x)=i^{m}\sqrt{|\det L|}\psi(Lx)\text{ \ , \ }\widehat
{V}_{P}\psi(x)=e^{\frac{i}{2\hbar}Px^{2}}\psi(x) \label{MLVP}%
\end{equation}
($m$ corresponds to a choice of argument for $\det L$). $\operatorname*{Mp}%
(n)$ is a faithful representation of the double covering group of
$\operatorname*{Sp}(n)$; the projection $\pi:\operatorname*{Mp}%
(n)\longrightarrow\operatorname*{Sp}(n)$ is determined by its action on the
generators:%
\begin{equation}
\pi(\widehat{M}_{L,m})=M_{L}\text{ , }\pi(\widehat{V}_{P})=%
\begin{pmatrix}
I & 0\\
-P & I
\end{pmatrix}
\text{ , }\pi(\widehat{J})=J \label{mlvp}%
\end{equation}
($M_{L}$ defined by formula (\ref{ml})).

\begin{description}
\item[W3] Let $\widehat{S}$ be any of the two metaplectic operators associated
with $S\in\operatorname*{Sp}(n)$. The following \textit{metaplectic covariance
}formula holds: \textit{ }
\begin{equation}
W(\widehat{S}\psi)(z)=W\psi(S^{-1}z). \label{metacobis}%
\end{equation}

\end{description}

\noindent As a particular case of (\ref{metacobis}) we have
\begin{equation}
W(F\psi)(z)=W(\widehat{J}\psi)(z)=W\psi(-Jz). \label{metaco}%
\end{equation}

Recall that the Heisenberg operator $\widehat{T}(z_{0})$ is defined, for
$z_{0}=(x_{0},p_{0})\in\mathbb{R}^{2n}$, by%
\begin{equation}
\widehat{T}(z_{0})\psi(x)=e^{\frac{i}{\hbar}(p_{0}\cdot x-\frac{1}{2}%
p_{0}\cdot x_{0})}\psi(x-x_{0}). \label{hw}%
\end{equation}

\begin{description}
\item[W4] We have%
\begin{equation}
W(\widehat{T}(z_{0})\psi)(z)=W\psi(z-z_{0}). \label{wt}%
\end{equation}

\end{description}

The Wigner transform behaves well under tensor products: if $x=(x^{\prime
},x^{\prime\prime})$ with $x^{\prime}\in\mathbb{R}^{k}$, $x^{\prime\prime}%
\in\mathbb{R}^{n-k}$ and $\psi^{\prime}\in L^{2}(\mathbb{R}^{k})$,
$\psi^{\prime\prime}\in L^{2}(\mathbb{R}^{n-k})$, then%
\begin{equation}
W(\psi^{\prime}\otimes\psi^{\prime\prime})=W^{\prime}\psi^{\prime}\otimes
W^{\prime\prime}\psi^{\prime\prime} \label{tensor1}%
\end{equation}
where $W^{\prime}$ and $W^{\prime\prime}$ are the Wigner transforms on
$L^{2}(\mathbb{R}^{k})$ and $L^{2}(\mathbb{R}^{n-k})$, respectively. More
generally, if $W,W^{\prime},W^{\prime}$ now denote cross-Wigner
distributions:
\begin{equation}
W(\psi^{\prime}\otimes\psi^{\prime\prime},\phi^{\prime}\otimes\phi
^{\prime\prime})=W^{\prime}(\psi^{\prime},\phi^{\prime})\otimes W^{\prime
\prime}(\psi^{\prime\prime},\phi^{\prime\prime}). \label{tensor2}%
\end{equation}

\subsection{Wigner transform and Hermite functions}

The $k$-th state of the quantum harmonic oscillator with classical Hamiltonian
$H(x,p)=\frac{1}{2}(x^{2}+p^{2})$ is the Hermite function%
\[
\psi_{k}(x)=h_{k}(\tfrac{1}{\sqrt{\hbar}}x)e^{-\tfrac{1}{2\hbar}x^{2}}%
\]
($h_{k}$ the $k$-th Hermite polynomial). One shows that
\begin{equation}
W(\psi_{k},\psi_{\ell})(z)=e^{-\tfrac{1}{\hbar}|z|^{2}}\sum_{j=0}^{\min
(k,\ell)}C_{j}(k,\ell)z^{\ell-j}\overline{z}^{k-j} \label{ha1}%
\end{equation}
where the $C_{j}(k,\ell)$ are real constants and $z$ is identified with
$x+ip\in\mathbb{C}^{n}$ in the right-hand side (see \textit{e.g.} \cite{fo89},
p. 66--67). Notice that in particular%
\begin{equation}
|W(\psi_{k},\psi_{\ell})(z)|\leq e^{-\tfrac{1}{\hbar}|z|^{2}}P_{k\ell}(|z|)
\label{ha2}%
\end{equation}
where $P_{k\ell}$ is a real polynomial of degree $k+\ell$.

We will need the following Lemma which says that the Wigner transform of a
Hermite function is the product of an exponential and of a polynomial with
positive leading coefficient. (For related results see \cite{jaei90}).

\begin{lemma}
\label{prop3}Let $Q$ be a (complex) polynomial function on $\mathbb{R}^{n}$
and $\psi(x)=Q(x)e^{-\frac{1}{2\hbar}Ax^{2}}$, $A>0$. Then:

(i) The Wigner transform of $\psi$ is given by
\begin{equation}
W\psi(x,p)=R(z_{A},\overline{z_{A}})e^{-\tfrac{1}{\hbar}|z_{A}|^{2}}
\label{for1}%
\end{equation}
where $R$ is a polynomial function and $z_{A}=A^{1/2}x+iA^{-1/2}p$ ($A^{1/2}$
the positive square root of $A$);

(ii) In particular%
\begin{equation}
|W\psi(x,p)|\leq T(|z_{A}|)e^{-\tfrac{1}{\hbar}|z_{A}|^{2}} \label{for2}%
\end{equation}
where $T$ is a polynomial with real coefficients.
\end{lemma}

\begin{proof}
(i) Let us set $\varphi=\widehat{M}_{A^{-1/2},0}\psi$ where $\widehat
{M}_{A^{-1/2},0}\in\operatorname*{Mp}(n)$ is defined by (\ref{MLVP}). Thus%
\[
\varphi(x)=P(x)e^{-\frac{1}{2\hbar}|x|^{2}}\text{ \ with \ }P(x)=\sqrt{\det
A^{-1}}Q(A^{-1/2}x)
\]
and we have, by property (\ref{metacobis}) of the Wigner transform and the
first formula (\ref{mlvp}),
\begin{equation}
W\psi(z)=W\varphi(A^{1/2}x,A^{-1/2}p). \label{wa}%
\end{equation}
Writing $P(x)=\sum_{\alpha}a_{\alpha}x^{\alpha}$ (we are using multi-index
notation $\alpha=(\alpha_{1},...,\alpha_{n})$, $x^{\alpha}=x_{1}^{\alpha_{1}%
}\cdot\cdot\cdot x_{n}^{\alpha_{n}}$) we have%
\[
\varphi(x)=\sum\nolimits_{\alpha}a_{\alpha}\varphi^{\alpha}(x)\text{ \ ,
\ }\varphi^{\alpha}=\varphi_{1}^{\alpha_{1}}\otimes\cdot\cdot\cdot
\otimes\varphi_{n}^{\alpha_{n}}%
\]
with $\varphi_{j}^{\alpha_{j}}(x_{j})=x_{j}^{\alpha_{j}}e^{-x_{j}^{2}/2}$. By
the sesquilinearity of the cross-Wigner transform we get
\begin{equation}
W\varphi=\sum\nolimits_{\alpha,\beta}a_{\alpha}\overline{a_{\beta}}%
W(\varphi^{\alpha},\varphi^{\beta}) \label{wifi}%
\end{equation}
and by the tensor product property (\ref{tensor2})%
\begin{align*}
W(\varphi^{\alpha},\varphi^{\beta})  &  =W(\varphi_{1}^{\alpha_{1}}%
\otimes\cdot\cdot\cdot\otimes\varphi_{n}^{\alpha_{n}},\varphi_{1}^{\beta_{1}%
}\otimes\cdot\cdot\cdot\otimes\varphi_{n}^{\beta_{n}})\\
&  =W(\varphi_{1}^{\alpha_{1}},\varphi_{1}^{\beta_{1}})\otimes\cdot\cdot
\cdot\otimes W(\varphi_{n}^{\alpha_{n}},\varphi_{n}^{\beta_{n}}).
\end{align*}
The Hermite functions $\psi_{k}$ forming an orthonormal basis of
$L^{2}(\mathbb{R})$ each $\varphi_{j}^{\alpha_{j}}$ is a finite linear
combination of these functions; using again sesquilinearity and applying
formula\ (\ref{ha1}) there exist polynomials $P_{\alpha_{j},\beta_{j}}$ such
that
\[
W(\varphi_{j}^{\alpha_{j}},\varphi_{j}^{\beta_{j}})(x_{j},p_{j})=P_{\alpha
_{j},\beta_{j}}(z_{j},\overline{z_{j}})e^{-\tfrac{1}{\hbar}|z_{j}|^{2}}%
\]
with $z_{j}=x_{j}+ip_{j}$ and hence%
\[
W(\varphi^{\alpha},\varphi^{\beta})(z)=P_{\alpha\beta}(z,\overline
{z})e^{-\tfrac{1}{\hbar}|z|^{2}}%
\]
where $P_{\alpha\beta}=P_{\alpha_{1},\beta_{1}}\otimes\cdot\cdot\cdot\otimes
P_{\alpha_{n},\beta_{n}}$ is a polynomial function in $2n$ variables. It
follows from (\ref{wifi}) that
\begin{equation}
W\varphi(z)=\sum\nolimits_{\alpha,\beta}a_{\alpha}\overline{a_{\beta}%
}P_{\alpha\beta}(z,\overline{z})e^{-\tfrac{1}{\hbar}|z|^{2}}=R(z,\overline
{z})e^{-\tfrac{1}{\hbar}|z|^{2}} \label{aap}%
\end{equation}
and hence, in view of (\ref{wa}),%
\[
W\psi(z)=\sum\nolimits_{\alpha,\beta}a_{\alpha}\overline{a_{\beta}}%
P_{\alpha\beta}(A^{1/2}x,A^{-1/2}p)e^{-\tfrac{1}{\hbar}(A^{-1}x^{2}+Ap^{2})}%
\]
as claimed. (ii) Since $W\varphi$ is a real function we have $W\varphi
(z)\leq|W\varphi|$ and hence, taking (\ref{aap}) into account,%
\[
W\varphi(z)\leq|R(z,\overline{z})|e^{-\tfrac{1}{\hbar}|z|^{2}}\leq
T(|z|)e^{-\tfrac{1}{\hbar}|z|^{2}}%
\]
which concludes the proof in view of (\ref{wa}).
\end{proof}

\subsection{Phase-space formulation of Hardy's theorem\label{sec4}}

When dealing with Gaussian functions related to \textquotedblleft squeezed
coherent states\textquotedblright\ we obtain Gaussian estimates where the
quadratic form in the exponent no longer is block-diagonal. For instance, the
Wigner transform of a Gaussian of the type
\[
\psi_{X,Y}(x)=e^{-\tfrac{1}{2\hbar}(X+iY)x^{2}}%
\]
($X$ and $Y$ real symmetric, $X>0$) is given by the formula
\begin{equation}
W\psi_{X,Y}(z)=(\pi\hbar)^{-n/2}(\det X)^{-1/2}e^{-\tfrac{1}{2\hbar}Gz^{2}}
\label{a1}%
\end{equation}
where the matrix $G$ is given by%
\begin{equation}
G=%
\begin{pmatrix}
X+YX^{-1}Y & YX^{-1}\\
X^{-1}Y & X^{-1}%
\end{pmatrix}
\label{a2}%
\end{equation}
(see Proposition 8.4, p. 263, in de Gosson \cite{Birk}); the result seems to
go back to Bastiaans according to Littlejohn \cite{Littlejohn}). An important
observation is that $G$ is a positive-definite symplectic matrix as follows
from the obvious factorization%
\begin{equation}
G=S^{T}S\text{ \ \textit{with} \ }S=%
\begin{pmatrix}
X^{1/2} & 0\\
X^{-1/2}Y & X^{-1/2}%
\end{pmatrix}
\in\operatorname*{Sp}(n)\text{.} \label{a3}%
\end{equation}
Setting $\Sigma=\frac{\hbar}{2}G^{-1}$ the ellipsoid $\mathcal{W}:\frac{1}%
{2}\Sigma^{-1}z^{2}\leq1$ is the set $\{z:S^{T}Sz^{2}\leq\hbar\}$;
$\mathcal{W}$ is thus the image of the ball $B(\sqrt{\hbar})$ by a linear
symplectic transformation, and thus has symplectic capacity $\frac{1}{2}h$.

Let us now show, as claimed in the introduction, that Hardy's uncertainty
principle for a function $\psi$ is equivalent to a condition on its Wigner
transform $W\psi$.

\begin{theorem}
\label{EquivalenceUP}\label{propmain}Let $\psi\in L^{2}(\mathbb{R}^{n})$ and
$A,B$ two positive real $n\times n$ matrices. Let $C_{A},C_{B}>0$. The
condition
\begin{equation}
|\psi(x)|\leq C_{A}e^{-\tfrac{1}{2\hbar}Ax^{2}}\text{ \ and \ }|F\psi(p)|\leq
C_{B}e^{-\tfrac{1}{2\hbar}Bp^{2}}\text{ } \label{cond1}%
\end{equation}
is equivalent to the existence of a constant $C_{AB}>0$ such that
\begin{equation}
W\psi(z)\leq C_{AB}e^{-\tfrac{1}{\hbar}(Ax^{2}+Bp^{2})}. \label{cond2}%
\end{equation}

\end{theorem}

\begin{proof}
In view of properties (\ref{wpsi2}) of $W\psi$, condition (\ref{cond2})
implies that there exist constants $C_{A},C_{B}\geq0$ such that
\[
|\psi(x)|^{2}\leq C_{A}^{2}e^{-\tfrac{1}{\hbar}Ax^{2}}\text{ \ \textit{,}
\ }|F\psi(p)|^{2}\leq C_{B}^{2}e^{-\tfrac{1}{\hbar}Bp^{2}}%
\]
hence (\ref{cond2})$\Longrightarrow$(\ref{cond1}). Let us prove that
conversely (\ref{cond1})$\Longrightarrow$(\ref{cond2}). Let $\lambda
_{1},...,\lambda_{n}$ be the eigenvalues of $AB$. If there exists $j$ such
that $\lambda_{j}>1$ then $\psi=0$ by Theorem \ref{prop1} and (\ref{cond2}) is
trivially verified. We may thus assume from now on that $\lambda_{j}\leq1$ for
$j=1,...,n.$ Let $L\in GL(n,\mathbb{R})$ be as in Lemma \ref{lem1}, that is
$L^{T}AL=L^{-1}B(L^{T})^{-1}=\Lambda$ where $\Lambda$ is the diagonal matrix
whose eigenvalues are the $\sqrt{\lambda_{j}}$. We have, setting $\psi
_{L}(x)=\psi(Lx)$ as in the proof of Theorem \ref{prop1},%
\[
|\psi_{L}(x)|\leq C_{A}e^{-\tfrac{1}{2\hbar}\Lambda x^{2}}\text{ \ and
\ }|F\psi_{L}(p)|\leq C_{B}e^{-\tfrac{1}{2\hbar}\Lambda p^{2}}.
\]
Since $\lambda_{j}\leq1$ for all $j=1,...,n$ Theorem \ref{prop1} implies that
we have%
\[
\psi_{L}(x)=Q_{L}(x)e^{-\tfrac{1}{\hbar}\Lambda x^{2}}%
\]
where $Q_{L}$ is a polynomial function which is constant when all the
$\lambda_{j}$ are equal to one. It follows, by Lemma \ref{prop3} that%
\[
W\psi_{L}(z)\leq R_{L}(y_{1}...,y_{n})e^{-\tfrac{1}{\hbar}(\Lambda
x^{2}+\Lambda^{-1}p^{2})}%
\]
where $R_{L}$ is a polynomial function with positive leading coefficient and
$y_{j}=\lambda_{j}x_{j}^{2}+\lambda_{j}^{-1}p_{j}^{2}$. Let $C_{L}>0$ be a
constant such that%
\[
W\psi_{L}(z)\leq C_{L}\prod\limits_{j=1}^{n}y_{j}^{m_{j}}e^{-\tfrac{1}{\hbar
}(\Lambda x^{2}+\Lambda^{-1}p^{2})};
\]
for every $\varepsilon>0$ there exists $C_{L,\varepsilon}>0$ such that%
\[
\prod\limits_{j=1}^{n}y_{j}^{m_{j}}e^{-\tfrac{1}{\hbar}(\Lambda x^{2}%
+\Lambda^{-1}p^{2})}\leq C_{L,\varepsilon}e^{-\tfrac{1}{\hbar}((\Lambda
-\varepsilon)x^{2}+\Lambda^{-1}p^{2})}%
\]
(we are writing $\Lambda-\varepsilon$ for $\Lambda-I\varepsilon$) and hence%
\begin{equation}
W\psi_{L}(z)\leq C_{L,\varepsilon}e^{-\tfrac{1}{\hbar}((\Lambda-\varepsilon
)x^{2}+\Lambda^{-1}p^{2})}. \label{w1}%
\end{equation}
Applying the same argument to $W(F\psi_{L})(x,p)=W\psi_{L}(-p,x)$ we also have%
\begin{equation}
W\psi_{L}(z)\leq C_{L,\varepsilon}e^{-\tfrac{1}{\hbar}(\Lambda x^{2}%
+(\Lambda^{-1}-\varepsilon)p^{2})}. \label{w2}%
\end{equation}
Since%
\[
\sup[(\Lambda-\varepsilon)x^{2}+\Lambda^{-1}p^{2},\Lambda x^{2}+(\Lambda
^{-1}-\varepsilon)p^{2}]=\Lambda x^{2}+\Lambda^{-1}p^{2}%
\]
the inequalities (\ref{w1})--(\ref{w2}) imply that we have%
\[
W\psi_{L}(z)\leq C_{L,\varepsilon}e^{-\tfrac{1}{\hbar}(\Lambda x^{2}%
+\Lambda^{-1}p^{2})}\text{;}%
\]
since $\psi_{L}(x)=\psi(Lx)$ this is just condition (\ref{cond2}).
\end{proof}

Theorem \ref{propmain} has the following consequence which contains Hardy's
theorem as a particular case (we have proved a particular case of that result,
using different methods, in \cite{GosLue,Goluahp}).

\begin{corollary}
\label{comain}Let $\psi\in L^{2}(\mathbb{R}^{n}),$ $\psi\neq0$. Assume that
there exists a positive-definite real matrix $M$, a vector $a\in
\mathbb{R}^{2n}$ and $C>0$ such that%
\begin{equation}
W\psi(z)\leq Ce^{-\tfrac{1}{\hbar}(Mz^{2}+2a\cdot z)}. \label{wpsim}%
\end{equation}
Then the ellipsoid $\mathcal{W}=\{z:Mz^{2}\leq\hbar\}$ has symplectic capacity
$c(\mathcal{W})\geq\frac{1}{2}h$ (equivalently $\lambda_{1}^{\sigma}\leq\hbar$).
\end{corollary}

\begin{proof}
Assume first that $a=0$. Let $S\in\operatorname*{Sp}(n)$ be such that
$S^{T}MS$ is in Williamson diagonal form
\[
D=%
\begin{pmatrix}
\Lambda & 0\\
0 & \Lambda
\end{pmatrix}
\]
with $\Lambda=\operatorname*{diag}(\lambda_{1}^{\sigma},...,\lambda
_{n}^{\sigma})$, $\lambda_{1}^{\sigma}\geq\cdot\cdot\cdot\geq\lambda
_{n}^{\sigma}$. Choose $\widehat{S}\in\operatorname*{Mp}(n)$ with projection
$S$. It follows from the metaplectic covariance property (\ref{metacobis}) of
the Wigner transform that%
\[
W(\widehat{S}^{-1}\psi)(z)\leq Ce^{-\tfrac{1}{\hbar}(\Lambda x^{2}+\Lambda
p^{2})}\text{.}%
\]
Applying Theorem (\ref{propmain}) there exist constants $C_{1},C_{2}>0$ such
that
\[
|\widehat{S}^{-1}\psi(x)|\leq C_{1}e^{-\tfrac{1}{2\hbar}\Lambda x^{2}}\text{
\ and \ }|F\widehat{S}^{-1}\psi(x)|\leq C_{2}e^{-\tfrac{1}{2\hbar}\Lambda
p^{2}}.
\]
In view of the multidimensional Hardy theorem \ref{prop1} we must have
$\lambda_{j}^{\sigma}\leq1$ for $j=1,...n$ hence (Lemma \ref{lem3})
$c(\mathcal{W)}=\pi\hbar/\lambda_{1}^{\sigma}\geq\pi\hbar$ which concludes the
proof in the case $a=0$. Assume now $a$ is arbitrary, and set $Q(z)=Mz^{2}%
+2a\cdot z$; choosing $S\in\operatorname*{Sp}(n)$ and $\widehat{S}%
\in\operatorname*{Mp}(n)$ as above we have $Q(Sz)=Dz^{2}+2b\cdot z$ where
$b=S^{T}a$; completing squares we get%
\[
Q(Sz)=D(z+D^{-1}b)^{2}-D^{-1}b^{2}.
\]
It follows that for a new constant $C^{\prime}$ we have
\[
W(\widehat{S}^{-1}\psi)(z)\leq Ce^{-\tfrac{1}{\hbar}Q(Sz)}\leq C^{\prime
}e^{-\tfrac{1}{\hbar}D(z+D^{-1}b)^{2}}.
\]
We next observe that
\[
W(\widehat{S}^{-1}\psi)(z-D^{-1}b)=W(\widehat{T}(b)\widehat{S}^{-1}\psi)(z)
\]
where $\widehat{T}(D^{-1}b)$ is a Heisenberg operator; we thus have, using
(\ref{wt}),%
\[
W(\widehat{T}(D^{-1}b)\widehat{S}^{-1}\psi)(z)\leq C^{\prime}e^{-\tfrac
{1}{\hbar}Dz^{2}}%
\]
and it now suffices to apply the case $a=0$ to $\psi^{\prime}=\widehat
{T}(D^{-1}b)\widehat{S}^{-1}\psi$.
\end{proof}

It is instructive to see how the sub-Gaussian estimate (\ref{wpsim}) is
related to the uncertainty principle of Quantum Mechanics. Setting
$\Sigma=\frac{\hbar}{2}M^{-1}$ we can define the multivariate Gaussian
probability density%
\[
\rho(z)=\left(  \frac{1}{2\pi}\right)  ^{n}(\det\Sigma)^{-1/2}e^{-\frac{1}%
{2}\Sigma^{-1}z^{2}}%
\]
and view $\Sigma$ as a statistical covariance matrix. The Wigner ellipsoid
$\mathcal{W}:Mz^{2}\leq\hbar$ is identical with the set $\mathcal{W}_{\Sigma
}=\{z:\tfrac{1}{2}\Sigma^{-1}z^{2}\leq1\}$.

One of us has proven in \cite{Birk} the following result:

\begin{proposition}
\label{propcw}The two following conditions are equivalent:

(i) $c(\mathcal{W})=c(\mathcal{W}_{\Sigma})\geq\frac{1}{2}h$

(ii) \textit{The Hermitian matrix }$\Sigma+i\frac{\hbar}{2}J$\textit{ is
positive semidefinite.}
\end{proposition}

Write now $\Sigma$ in block-matrix form%
\[
\Sigma=%
\begin{pmatrix}
\Sigma_{XX} & \Sigma_{XP}\\
\Sigma_{PX} & \Sigma_{PP}%
\end{pmatrix}
\]
where $\Sigma_{XX},$ $\Sigma_{XP}=\Sigma_{PX}^{T}$, and $\Sigma_{PP}$ are the
$n\times n$ partial covariance matrices $\Sigma_{XX}=(\operatorname*{Cov}%
(x_{j},x_{k}))_{j,k}$, $\Sigma_{XP}=\Sigma_{XP}^{T}=(\operatorname*{Cov}%
(x_{j},p_{k}))_{j,k}$, and $\Sigma_{PP}=(\operatorname*{Cov}(p_{j}%
,p_{k}))_{j,k}$; the covariances are defined with respect to the probability
density $\rho_{\Sigma}$: setting $z_{j}=x_{j}$ for $1\leq j\leq n$ and
$z_{j}=p_{j}$ for $n+1\leq j\leq2n$ we have%
\[
\operatorname*{Cov}(z_{j},z_{k})=\int z_{j}z_{k}\rho(z)d^{2n}z-\int z_{j}%
\rho(z)d^{2n}z\int z_{k}\rho(z)d^{2n}z
\]
The conditions (i) and (ii) in Proposition \ref{propcw} above are equivalent
to%
\begin{equation}
(\Delta x_{j})^{2}(\Delta p_{j})^{2}\geq(\operatorname*{Cov}(x_{j},p_{j}%
))^{2}+\tfrac{1}{4}\hbar^{2} \label{unc2}%
\end{equation}
for $j=1,...,n$ (see Narcowich \cite{na90} and de Gosson \cite{Birk}, and the
references therein). The inequalities (\ref{unc2}) (known in the
quantum-mechanical literature as the Schr\"{o}dinger--Robertson uncertainty
relations) are a precise form of the usual text-book Heisenberg inequalities
$\Delta X_{j}\Delta P_{j}\geq\tfrac{1}{2}\hbar$ to which they reduce if one
neglects correlations.

\section{Two Extensions of Hardy's Theorem\label{seclast}}

\subsection{Restatement in an arbitrary Lagrangian frame\label{seclast1}}

The statements of Hardy's uncertainty principle we have been considering
correspond to a particular choice of coordinates namely the positions $x$ and
the momenta $p$ for the phase space. These statements thus correspond to the
choice of frame $(\ell_{X},\ell_{P})$ where $\ell_{X}$ is the horizontal
Lagrangian plane $\mathbb{R}^{n}\times\{0\}$ and $\ell_{P}$ the vertical
Lagrangian plane $\{0\}\times\mathbb{R}^{n}$. This choice is of course to a
great extent arbitrary. In the following we are going to extend our results to
arbitrary Lagrangian frames.

Recall that a subspace $\ell$ of the phase space $(\mathbb{R}^{2n},\sigma)$ is
called \textit{isotropic}, if the symplectic form $\sigma$ vanishes
identically on $\ell$. If $\ell$ has maximal dimension $n$, then $\ell$ is
called a \textit{Lagrangian plane}. The set of all Lagrangian planes in
$(\mathbb{R}^{2n},\sigma)$ is called the \textit{Lagrangian Grassmannian} of
$(\mathbb{R}^{2n},\sigma)$ and denoted by $\operatorname{Lag}(n)$.

The subgroup of all symplectic matrices $S$ such that $S\ell=\ell$ is called
the \textit{stabilizer of }$\ell$ and denoted by $St(\ell)$. Note that $S\in
St(\ell)$ if and only if $S^{T}\in St(J\ell)$.

If $\ell$ and $\ell^{\prime}$ are Lagrangian planes in $(\mathbb{R}%
^{2n},\sigma)$ satisfying $\ell\cap\ell^{\prime}=0$, then $(\ell,\ell^{\prime
})$ are called \textit{transversal}; equivalently $\ell\oplus\ell^{\prime
}=\mathbb{R}^{2n}$. We will call a pair $(\ell,\ell^{\prime})$ of transversal
Lagrangian planes a \textit{Lagrangian frame}. An important property is that
the symplectic group $\operatorname*{Sp}(n)$ acts transitively not only on the
Lagrangian Grassmannian $\operatorname{Lag}(n)$, but also on the set of
Lagrangian frames: if $(\ell_{1},\ell_{1}^{\prime})$ and $(\ell_{2},\ell
_{2}^{\prime})$ are pairs of Lagrangian planes satisfying $\ell_{1}\cap
\ell_{1}^{\prime}=\ell_{2}\cap\ell_{2}^{\prime}=0$, then there exists
$S\in\operatorname*{Sp}(n)$ such that $(\ell_{2},\ell_{2}^{\prime})=(S\ell
_{1},S\ell_{1}^{\prime})$ (see de Gosson \cite{Birk}).

We interpret the marginal properties (\ref{wpsi2}) of Wigner's distribution in
terms of the horizontal and vertical Lagrangian plane $\ell_{X}$ and $\ell
_{P}$: if $\psi,F\psi\in L^{1}(\mathbb{R}^{n})\cap L^{2}(\mathbb{R}^{n})$ then
we can rewrite (\ref{wpsi2}) as
\[
\int_{\ell_{X}}W\psi(z)dz=|F\psi(p)|^{2}\text{ \ , \ }\int_{\ell_{P}}%
W\psi(z)dz=|\psi(x)|^{2}.
\]
Recall that the metaplectic covariance property (W5) of the Wigner
distribution tells us that if $\widehat{S}\in\operatorname*{Mp}(n)$ has
projection $S$ on $\operatorname*{Sp}(n)$ then $W(\widehat{S}\psi
)(z)=W\psi(S^{-1}z)$. Therefore, we get
\[
\int_{\ell_{X}}W(\widehat{S}\psi)(z)dz=\int_{\ell_{X}}W\psi(S^{-1}z)dz,
\]
or, equivalently,
\[
\int_{S\ell_{X}}W\psi(z)dz=|\widehat{S}\psi(z)|^{2}\text{ for }z\in S\ell
_{X}.
\]
Since $J\ell_{X}=\ell_{P}$ and $\widehat{J}=F$ we get the analogous results
for the Lagrangian plane $SJ$. Note that $(S\ell_{X},SJ\ell_{X})$ are a
transversal pair of Lagrangian planes. In other words the transitivity of the
symplectic group on $\operatorname{Lag}(n)$ allows to translate a statement
about $\ell_{X}$ and $\ell_{P}$ into a statement of another pair of Lagrangian
planes $(\ell,\ell^{\prime})$. One just has to choose the correct $S$ to go
from $(\ell_{X},\ell_{P})$ to $(\ell,\ell^{\prime})$. Then the statements
about $\psi$ and $F\psi$ translate into statements about $\widehat{S}\psi$ and
$\widehat{S}\circ\widehat{J}\psi$. Consequently, one of our main results,
Theorem \ref{EquivalenceUP}, remains valid in an arbitrary Lagrangian frame,
if one makes the proper modifications as indicate above.

\subsection{The case of convex exponents\label{seclast2}}

We are going to extend the Corollary \ref{comain} of Theorem \ref{propmain} to
the case where the inequality (\ref{wpsim}) is replaced by%
\begin{equation}
W\psi(z)\leq Ce^{-\tfrac{1}{\hbar}Q(z)} \label{wq}%
\end{equation}
where $Q$ is a uniformly convex function on $\mathbb{R}^{2n}$; we will assume
that $Q(0)=0$ (the case $Q(0)\neq0$ is trivially reduced to this case by
changing the constant $C$). Using the same trick as in the proof of Corollary
\ref{comain} we may moreover assume, replacing $\psi$ by $\widehat{T}%
(z_{0})\psi$ for a suitably chosen $z_{0}\in\mathbb{R}^{2n}$, that
\[
Q^{\prime}(0)=\nabla_{z}Q(0)=0.
\]

Let us briefly recall a few basic facts on convex functions (see Andrei
\cite{nec} for a concise review of the topic). A function $Q:\mathbb{R}%
^{2n}\longrightarrow\mathbb{R}$ is strictly convex if we have%
\[
Q(\alpha z+(1-\alpha)z^{\prime})<\alpha Q(z)+(1-\alpha)Q(z^{\prime})
\]
for $0<\alpha<1$ and $z\neq z^{\prime}$. If the function $Q$ is of class
$C^{2}$ this condition is equivalent to $Q^{\prime\prime}(z)>0$ for all
$z\in\mathbb{R}^{2n}$ ($Q^{\prime\prime}(z)$ is the Hessian matrix of $Q$
calculated at $z$). In what follows we will make in addition the following
uniformity assumption:%
\begin{equation}
\text{\textit{There exists}}\mathit{\ }c>0\mathit{\ }\text{\textit{such that}%
}\mathit{\ }Q^{\prime\prime}(z_{0})z^{2}\geq c|z|^{2}\mathit{\ }%
\text{\textit{for all}}\mathit{\ }z,z_{0}\in\mathbb{R}^{2n}\mathit{.}
\label{c1}%
\end{equation}
Diagonalizing $Q^{\prime\prime}(z_{0})$ by an orthogonal matrix $H(z_{0})$
this condition is easily seen to be equivalent to the condition $\lambda
_{Q}>0$ where%
\begin{equation}
\lambda_{Q}=\inf_{z_{0}\in\mathbb{R}^{2n}}\{\lambda(z):\lambda(z)\text{
\textit{is an eigenvalue of} }Q^{\prime\prime}(z)\}\text{;} \label{c2}%
\end{equation}
the smallest constant $c$ for which (\ref{c1}) holds is then precisely
$\lambda_{Q}$.

\begin{proposition}
\label{propq}Under the same assumptions on $Q$ as above let $\psi\in
L^{2}(\mathbb{R}^{2n})$, $\psi\neq0$ be such that $W\psi(z)\leq Ce^{-\tfrac
{1}{\hbar}Q(z)}$ for some $C>0$. Then the convex set
\begin{equation}
\mathcal{C}=\{z\in\mathbb{R}^{2n}:Q(z)\leq\hbar\} \label{coco}%
\end{equation}
satisfies $c(\mathcal{C})\geq\frac{1}{2}h$ for every symplectic capacity $c$
on $(\mathbb{R}^{2n},\sigma)$.
\end{proposition}

\begin{proof}
Let us begin by showing that we have $0<\lambda_{Q}\leq2$. Since the
uniformity assumption is equivalent to $\lambda_{Q}>0$ it suffices to show
that $\lambda_{Q}\leq2$. In view of the mean value theorem we have%
\begin{equation}
Q(z)=Q(0)+Q^{\prime}(0)\cdot z+\tfrac{1}{2}Q^{\prime\prime}(z^{\prime}%
)z^{2}=\tfrac{1}{2}Q^{\prime\prime}(z^{\prime})z^{2} \label{qo}%
\end{equation}
where $z^{\prime}$ lies on the line segment joining $0$ to $z$; we have
$Q^{\prime\prime}(z^{\prime})z^{2}\geq\lambda_{Q}|z|^{2}$ hence%
\begin{equation}
Q(z)\geq\tfrac{1}{2}\lambda_{Q}|z|^{2} \label{q1}%
\end{equation}
so that%
\[
W\psi(z)\leq Ce^{-\tfrac{1}{\hbar}Q(z)}\leq C^{\prime}e^{-\tfrac{1}{2\hbar
}\lambda_{Q}|z|^{2}}.
\]
The symplectic spectrum of $\lambda_{Q}I$ consists of the point $\lambda_{Q}$
hence we must have $\lambda_{Q}\leq2$ choosing $M=\frac{1}{2}\lambda_{Q}I$ in
Corollary \ref{comain}. The proposition follows: in view of (\ref{q1}) the
condition $Q(z)\leq\hbar$ implies $\tfrac{1}{2}\lambda_{Q}|z|^{2}\leq\hbar$
hence the set $\mathcal{C}$ contains the ball $|z|^{2}\leq2\hbar/\lambda
_{Q}\leq\hbar$; one concluded using the monotonicity of symplectic capacities
(Property (SC1)).
\end{proof}

Let $\mathcal{C}$ be a compact and convex set. We recall (John \cite{jo48})
that there exists a unique ellipsoid $\mathcal{W}$ contained in $\mathcal{C}$
having maximal volume. This ellipsoid, called the \textquotedblleft John
ellipsoid\textquotedblright\ \ref{fj}, has the property that%
\begin{equation}
\mathcal{W}\subset\mathcal{C}\subset z_{0}+2n(\mathcal{W}-z_{0}) \label{wc}%
\end{equation}
where $z_{0}$ is the center of $\mathcal{W}$. The result above has the
following immediate consequence:

\begin{corollary}
Let $\mathcal{W}$ be the John ellipsoid associated to the convex and compact
set (\ref{coco}). We have $c(\mathcal{W})\geq\frac{1}{2}h$.
\end{corollary}

\begin{proof}
The uniform convexity of $Q$ implies that the convex set $\mathcal{C}%
=\{z:Q(z)\leq\hbar\}$ is compact (Andrei \cite{nec}); John's ellipsoid is thus
well-defined. In the proof of Proposition \ref{propq} we have seen that
$\mathcal{C}$ contains the ball $|z|^{2}\leq2\hbar/\lambda_{Q}\leq\hbar$; this
ball is contained in John's ellipsoid. The result follows again in view of the
monotonicity of a symplectic capacity.
\end{proof}

Proposition \ref{propq} has another interesting non-trivial consequence. In
\cite{hoze94} Hofer and Zehnder construct a symplectic capacity $c_{\text{HZ}%
}$ having the following property:

\begin{quotation}
\textit{If }$\Omega$\textit{ is a convex and compact subset of }$R^{2n}%
$\textit{ with smooth boundary }$\partial\Omega$\textit{ then }%
\[
c_{\text{HZ}}(\Omega)=\inf_{\gamma}\left\{  \oint\nolimits_{\gamma
}pdx\right\}
\]
\textit{where }$\gamma$\textit{ ranges over the set of all periodic
Hamiltonian orbits on }$\partial\Omega$\textit{.}
\end{quotation}

\begin{corollary}
Under the same assumptions on $Q$ and $\psi$ as above we have
\begin{equation}
\oint\nolimits_{\gamma}pdx\geq\frac{1}{2}h \label{pdx}%
\end{equation}
for every periodic Hamiltonian orbit $\gamma$ on the hypersurface defined by
$Q(z)=\hbar$.
\end{corollary}

\begin{proof}
The boundary of $\mathcal{C}$ is precisely the hypersurface defined by
$Q(z)=\hbar$. In view of Proposition \ref{propq} we have $c(\mathcal{C}%
)\geq\frac{1}{2}h$ for every symplectic capacity $c$ hence, choosing
$c=c_{\text{HZ}}$
\[
\inf_{\gamma}\left\{  \oint\nolimits_{\gamma}pdx\right\}  \geq\frac{1}{2}h
\]
which proves (\ref{pdx}).
\end{proof}

\section{Concluding Remarks}

We have proved a $n$-dimensional version of Hardy's uncertainty principle, and
showed that it is equivalent to a statement on the Wigner distribution of a
sub-Gaussian state. The extension of this result to more general estimates
involving convex exponents in Subsection \ref{seclast2} opens the door to the
study of non-trivial properties for the density matrix of quantum systems.
Such applications are very important for the understanding of non-linear
quantum optics and the theory of entangled quantum states.

We mention that Hogan and Lakey \cite{hola05} have done a very interesting
analysis of the interplay between Hardy's uncertainty principle and rotations.
It would certainly be useful to restate their results in our context; we leave
this possibility for further work. Also, Gr\"{o}chenig and Zimmermann
\cite{grzi01} have studied Gaussian estimates from the point of view of the
short-time Fourier transforms; the methods they use are very different from ours.

\textit{E-mail : maurice.de.gosson@univie.ac.at (M. de Gosson)}

\textit{E-mail : franz.luef@univie.ac.at (F. Luef)}

\end{document}